\documentclass[a4paper,11pt]{article}
\pdfoutput=1 
\usepackage{jheppub,xcolor} 

\usepackage[T1]{fontenc} 

\newcommand{\cN}{\mathcal{N}}
\newcommand{\pd}{\partial}
\newcommand{\ol}[1]{\overline{#1}}

\newcommand{\cV}{\mathcal{V}}

\newcommand{\vp}{\varphi}

\newcommand{\cG}{\mathcal{G}}

\def\K{K{\"a}hler}
\def\be{\begin{equation}}
\def\ee{\end{equation}}
\newcommand{\ba}{\begin{eqnarray}}
\newcommand{\ea}{\end{eqnarray}}

\newcommand{\rf}[1]{(\ref{#1})}

\colorlet{darkgreen}{green!50!black}
\colorlet{violetgreen}{green!40!purple}
\title{\rm \bf \Huge Fibre Inflation and {\boldmath {$\alpha$}-attractors}}

\author[a,b]{Renata Kallosh,}
\author[a,b]{Andrei Linde,}
\author[c]{Diederik Roest,}
\author[d]{Alexander Westphal}
\author[a]{and Yusuke Yamada}
\affiliation[a]{Stanford Institute for Theoretical Physics and Department of Physics, Stanford University, Stanford, CA 94305, USA}
\affiliation[b]{Lorentz Institute for theoretical physics, University of Leiden, 2333CA Leiden, The Netherlands}
\affiliation[c]{Van Swinderen Institute for Particle Physics and Gravity, 
University of Groningen, Nijenborgh 4, 9747 AG Groningen, The Netherlands}
\affiliation[d]{Deutsches Elektronen-Synchrotron DESY, Theory Group, D-22603 Hamburg, Germany}
\emailAdd{kallosh@stanford.edu}
\emailAdd{alinde@stanford.edu}
\emailAdd{d.roest@rug.nl}
\emailAdd{alexander.westphal@desy.de}
\emailAdd{yusukeyy@stanford.edu}

\preprint{DESY-17-105}

\abstract{Fibre inflation is a specific string theory construction based on the  Large Volume Scenario that produces an inflationary plateau. We outline its relation to $\alpha$-attractor models for inflation, with the cosmological sector originating from certain string theory corrections leading to $\alpha=2$ and $\alpha=1/2$. Above a certain field range, the steepening effect of higher-order corrections leads first to the breakdown of single-field slow-roll and after that to the onset of 2-field dynamics: the overall volume of the extra dimensions starts to participate in the effective dynamics.
Finally, we propose effective supergravity models of fibre inflation based on an ${\overline {D3}}$ uplift term with a nilpotent superfield. Specific moduli dependent $\overline {D3}$ induced geometries lead to cosmological  fibre models but have in addition a de Sitter minimum exit. These supergravity models motivated by fibre inflation are relatively simple, stabilize the axions and disentangle the Hubble parameter from supersymmetry breaking.}

\begin{document} 
\maketitle
\flushbottom

\section{Introduction}

 Inflation has since long held the promise of providing an observational window on physics of very high energy scales, and might even offer a glimpse of string theory. With the beautiful CMB measurements of Planck in hand \cite{Planck2013, Planck2015}, it is natural to wonder about the relation between models compatible with the data and possible string inflationary set-ups.

 Starting with the former, $\alpha$-attractors are a rather minimal and elegant class of bottom-up supergravity models, that match the current CMB data with $n_s = 1 - 2/N$ and predict $r = 12 \alpha / N^2$ in terms of the number of e-folds $N$ \cite{Kallosh:2013yoa}. These models can be understood as pole inflation models: as a single-field model, the kinetic function of the inflaton consists of a second order pole whose location we can choose to be at $\phi=0$. At the same time, the scalar potential in this non-canonical frame is an arbitrary but regular function, which is positive around the pole \cite{Galante:2014ifa}. Canonical normalization of the inflaton then leads to infinite stretching of the scalar potential near $\phi = 0$ in an exponentially approached semi-infinite plateau.

The only relevant parameter for this class of models is the curvature of the hyperbolic moduli space, set by $\alpha$ \cite{Carrasco:2015uma}. While this is a tunable parameter in $\cN = 1$ supergravity, it is generically not in string theory set-ups. Instead, one typically obtains a number of copies of hyperbolic spaces. A natural question regards the possible values of $\alpha$ that can be obtained by the interplay between the different moduli spaces.

 This interplay is illustrated by the recent M-theory/string theory/maximal supergravity inspired models based on seven hyperbolic disks geometries  \cite{Ferrara:2016fwe,Kallosh:2017ced,Kallosh:2017wnt}.
 These correspond to either a particular $G_2$ compactification from 11D to 4D, or a  toroidal reduction of string theory, or on $E_{7(7)} (\mathbb{R}) \supset [SL(2, \mathbb{R})]^7$symmetry of $\cN=8$ 4D supergravity.  A subsequent set of simple cosmological disk merger models was proposed in 
 \cite{Kallosh:2017ced,Kallosh:2017wnt}
with some constraints on  the moduli of the seven unit-size-disks, which lead to $\alpha$-attractor  models with  $3 \alpha = 1,2,3,4,5,6,7$. Some of these constraints required that 
 $T_i = T_j$. 
 
A natural generalization involves more general identifications between tori. The first example going beyond the simple identification above is $T_i = T_j^p$ with $p\neq 0,1$.   In this paper we will analyze the consequences of such an identification for the case of two moduli and $p=\pm 2$ (both sign choices being related by moduli inversion). Moreover we point out that this is equivalent to volume stabilization in Calabi-Yau compactifications of string theory, as performed explicitly in e.g.~the Large Volume Stabilization (LVS) scenario \cite{Balasubramanian:2005zx}.

The model class of string inflation setup coming closest to this is ``fibre inflation'' \cite{Cicoli:2008gp} and various followups, see e.g.~\cite{Burgess:2016owb}. Fibre inflation builds on LVS with a  ``fibre volume modulus'', providing the inflationary direction. Various string corrections produce an effective 4D kinetic term and scalar potential that shows at leading order the structure of pole inflation. We will outline and explain the possible $\alpha$-attractors that can arise in such a setting of a fibred Calabi-Yau compactification.

However, fibre inflation can also come with corrections to the kinetic function and scalar potential arising from string loop corrections ~\cite{vonGersdorff:2005bf,Berg:2005ja, Berg:2007wt} and/or higher superspace-derivative corrections~\cite{Ciupke:2015msa} (in the spirit of the generalized pole inflation paper \cite{Broy:2015qna}). Such corrections {might}  spoil the infinite plateau  {and  instead could produce} rising exponential corrections after a finite $O(10 M_p)$ plateau. While the higher superspace-derivative corrections are given in terms of a topological quantity of the underlying compactification~\cite{Ciupke:2015msa}, the string loop corrections~\cite{vonGersdorff:2005bf,Berg:2005ja, Berg:2007wt} produce two terms in the scalar potential arising from KK-modes of the two 4-cycles of a fibred Calabi-Yau and a third term arising from winding modes of strings wrapping the intersection between the two 4-cycles. We will discuss the argument for the existence of singular terms in the scalar potential for non-canonically normalized inflaton (from string loops and $\alpha'$ corrections), and we will argue that the proposed singular terms of \cite{Cicoli:2008gp, Burgess:2016owb}  are not necessarily present. Adressing the same issue in \cite{Broy:2015zba, Cicoli:2016chb} where the extra $(\alpha')^3$ corrections from \cite{Ciupke:2015msa} is interesting and requires an independent  analysis. 

Finally, a crucial ingredient of the large volume scenario, on which fibre inflation builds, is the uplift from the non-SUSY AdS to a Minkowski or a de Sitter minimum. The introduction of a nilpotent multiplet can easily accommodate this uplifting. When the choice of the \K\ frame for the disk geometry is given in a form suggested in \cite{Carrasco:2015uma,Kallosh:2017ced} with an inflaton shift symmetry, the superpotential or $S$-field metric break this symmetry. The inflationary dynamics can be introduced either 
via a simple contribution  to  the superpotential  \cite{Kallosh:2017ced} 
or to the $S$-field metric \cite{McDonough:2016der}. We will use here the  $\overline{D3}$ induced geometric inflation construction  based on  \K\ function as proposed in \cite{Kallosh:2017wnt}, where this method was shown to be efficient in the context of the disk merger 
cosmological models.

We will provide here full supergravity effective descriptions of the interplay between the nilpotent multiplet and the fibre modulus in a concrete supergravity model that captures the essential ingredients of fibre inflation.

\section{Fibre inflation}

\subsection{Volume stabilization}

Fibre inflation comprises a class of possible string theory models that rely on the existence of a fibre modulus in the Calabi-Yau compactification. In order to stabilize the overall volume, they rely on the large volume stabilization (LVS) mechanism. This requires the volume to be dominated by a single term, while also including at least one blow-up mode. An explicit fibre example is provided by the case of $\mathbb{C}P^4_{[1,1,2,2,6]}  [12]$ model with 
\begin{equation}\label{Vol_fibre}
\mathcal V=\lambda\left(\sqrt{\tau_1}\tau_2 - \gamma \tau_3^{3/2} \right) \,,
\end{equation}
where $\tau_1$ is associated with the volume of the $K3$-fibre, $\tau_2$ controls the overall volume and $\tau_3$ denotes the blow-up and $\beta, \gamma$ are constants.
Note that the \K\ potential is a homogeneous function of weight $3/2$, resulting in the absence of a scalar potential for $\cV$ at tree-level: this is the no-scale structure of Calabi-Yau compactifications. Therefore the volume is a flat direction at tree-level.

However, both the total volume as well as the blow-up mode can be stabilized by the inclusion of perturbative $\alpha'$-corrections to the \K\ potential, and non-perturbative corrections to the superpotential:
 \begin{align}
   K = -2 \log( \cV + \xi) \,, \quad W = W_0 + A_3 \exp( - a_3 T_3 ) \,, \label{K}
 \end{align}
 with $T_i  = \tau_i + i \chi_i$ the holomorphic versions of the four-cycle volumes $\tau_i$. The resuling potential reads
 \begin{align}
V=\frac{8a^2A^2}{3\gamma}\left(\frac{\sqrt{\tau_3}}{\cV}\right)e^{-2a\tau_3}-4W_0aA\left(\frac{\tau_3}{\cV^2}\right)e^{-a\tau_3}+\frac{3\xi W_0^2}{4\cV^3}.
\end{align}
This produces a minimum for $\tau_3$ and $\cV$ at exponentially large values of the latter: in the limit $a\tau_3\gg 1$ an analytic approximation is
 \begin{align}
    \cV =\frac{3 \gamma  \sqrt{\tau_3} {W_0} e^{a \tau_3}}{4 a A} \,, \quad \tau_3 =\left (\frac{\xi }{2\gamma }\right)^{2/3} \,, \quad \chi_3 = 0 \quad{\rm where:}\;\;\xi\sim -g_s^{-3/2}\chi_{CY} \label{LVS}
 \end{align}
and $\chi_{CY}$ denotes the Euler characteristic of the Calabi-Yau manifold.
This produces the well-known non-SUSY anti-de Sitter minimum of the LVS scenario, which is stabilized by a barrier that scales as $\cV^{-3}$. 

\subsection{Kinetic terms and pole inflation structure}

Finding $\alpha$-attractor-like regimes of pole inflation in a type IIB LVS compactification on a fibred CY requires finding volume moduli with 2nd order poles in the kinetic terms without corresponding poles in the scalar potential. In order to exhibit the pole structure of the two volume moduli of a fibred CY with LVS stabilization we need to include the kinetic terms of both moduli in \eqref{K}
 \begin{align}
  &{\cal L}_{kin.}= \sum_{i=1,2} -\frac{3\alpha_i}{4} \frac{\pd\tau_i\pd \tau_i}{\tau_i^2} =  -\frac14 \frac{\pd\tau_1\pd \tau_1}{\tau_1^2}- \frac12 \frac{\pd\tau_2\pd\tau_2}{\tau_2^2} \,. \label{kinetic}
 \end{align}
 Here, we focus on the real parts and  ignore axions for the moment.
 If we now impose volume stabilization a la LVS enforcing $\cV\simeq \lambda\sqrt{\tau_1}\tau_2\equiv \langle\cV\rangle=const.$ we are justified in dropping derivatives of the volume when we replace either $\tau_1$ or $\tau_2$ in terms of the other modulus. Hence, up to derivatives of the volume these two kinetic terms combine into
  \begin{align}
  &{\cal L}_{kin.} \simeq -\frac38 \frac{\pd\tau_1\pd \tau_1}{\tau_1^2} \simeq  -\frac32 \frac{\pd\tau_2\pd\tau_2}{\tau_2^2} \,. \label{kinetic2}
 \end{align}
 Thus, we get the relation 
\begin{align}
 \tau_2 = e^{- \varphi / \sqrt{3}} 
\end{align} 
for the effective canonically normalized inflaton field $\varphi$.


\subsection{Loop corrections}
  
In case the Calabi-Yau manifold is fibered, as in the example \eqref{Vol_fibre}, the leading volume term is a product. Stabilization of the overall volume therefore leaves a flat direction and hence provides a possible avenue for inflation. To produce a scalar potential with a minimum for the fibre modulus, one has to include further corrections. These can include a series of conjectured loop corrections of the form:
 \begin{align}
   \delta K = \frac{C^{KK}_{1}}{\tau_1} + \frac{C^{KK}_2}{\tau_2} + \frac{C^W_{12}}{\tau_1 \tau_2} \,, \label{loops}
 \end{align}
where the first two arise from the exchange of Kaluza-Klein (KK) modes, for example, between D7-branes and D3-branes, which are usually needed for tadpole cancellation. These corrections are suppressed by the volume of the 4-cycle wrapped by the D7-branes. In contrast, the third correction comes from the exchange of winding strings between intersecting stacks of D7-branes. All these terms have been calculated to exist in toroidal compactifications \cite{vonGersdorff:2005bf,Berg:2005ja}, and it has been argued that they should persist for Calabi-Yau generalizations \cite{Cicoli:2008gp}. Moreover, the coefficients $C^{KK}_i$ and $C^W_{12}$ are functions that depend on complex structure moduli $U$ which are stabilized at tree-level by background fluxes. As a consequence, the coefficients are assumed to be $O(1)$ constants. An important point of this expansion is that its consistency requires both $\tau_1$ and $\tau_2$ to be large. However, at fixed volume \eqref{Vol_fibre}, these two moduli are inversely proportional 
and hence this implies that there is a bound to the regime where these can be trusted. We will get back to this point later.

The above \K\ string loop corrections result in a scalar potential that is of the form
{\begin{align}\label{string_loops}
\delta V 
& = g_s\frac{|W_0|^2}{\mathcal V^2} \left( \frac{(g_s C_1^{KK})^2}{\tau_1^2} + 2 \frac{(g_s C_2^{KK})^2}{\tau_2^2} 
- \frac{2 C_{12}^W}{\lambda\tau_1 \tau_2} \right) + \delta_{\rm up} \quad . \notag \\
\end{align}
Here, an explicit uplift term has also been included in order to have viable inflation.
Upon LVS volume stabilization on a fibred CY we need to impose $\cV = \lambda \sqrt{\tau_1}\tau_2=const.$ on the previous expression. Then, we get
{\begin{align}\label{string_loops_v2}
\delta V & = g_s\frac{|W_0|^2}{\mathcal V^2} \left( \frac{(\lambda^2 g_s C_1^{KK})^2 \tau_2^4}{\cV^4} + 2 \frac{(g_s C_2^{KK})^2}{\tau_2^2} 
- \frac{2 \lambda C_{12}^W \tau_2}{\cV^2} \right)  + \delta_{\rm up} \,,
\end{align}
Note that the KK corrections to the \K\ potential drop out at leading order: this has been dubbed extended no-scale structure~\cite{Cicoli:2007xp}. 

We will now review the generic properties of the string loop corrections.


\begin{itemize}

\item The string loop corrections to the K\"ahler potential of a fibred 2-moduli Calabi-Yau manifold contain two contributions arising from KK-modes on 4-cycles wrapped by D7-branes which \emph{only intersect themselves}, and a third contribution arising from winding modes on a 1-cycle in the intersection of two 4-cycles which are \emph{both} wrapped by D7-branes (see the discussion in~\cite{Berg:2007wt,Cicoli:2016chb}). 

\item In general, there will be other smooth and connected 4-cycles required to be present due to D7-brane tadpole cancellation in a full CY orientifold model which intersect either $\tau_1$ or $\tau_2$ or each of them. D7-branes wrapped on those 4-cycles wrap the intersections with $\tau_1$ and/or $\tau_2$ as well. This will generate winding mode corrections even we only wrap either $\tau_1$ or $\tau_2$ but not both. Therefore, generically the winding mode corrections are expected to be present~\cite{Cicoli:2016xae}.

\item Similarly, a full 4D ${\cal N}=1$ CY orientifold model will in general contain O7, and O3 planes, as well as D3-branes. Additional KK mode corrections may then arise from the exchange of KK modes with these additional objects~\cite{Cicoli:2016xae}. We should therefore expect KK mode corrections of the form displayed in eq.~\eqref{string_loops} to be generically present.

\item Finally, we note here that all the above conclusions about the generic presence of all of the types of string loop corrections to $K$ rest on the extrapolation of the explicit toroidal orientifold calculations to the general CY case, which were performed in absence of any moduli stabilization scheme imposing a constraint like $\tau_1\sim 1/ \tau_2^2$ here. Hence, the correction terms were originally functions of $1/\tau_1$ and $1/\tau_2$ separately. If this form survives in presence of constraint relations between the moduli imposed by moduli stabilization, then all of the above conclusions about the presence and form of the string loop corrections follow. Therefore it would be important to check this conjecture with explicit string loop computations for CY moduli in the presence of volume stabilization mechanisms.

\end{itemize}

\subsection{Higher superspace-derivative corrections}

In addition to loop corrections, higher derivative corrections will also induce a potential for the initially flat fibre direction. These were calculated in \cite{Ciupke:2015msa} and subsequently employed for inflation in \cite{Broy:2015zba}. They are proportional to  integer numbers $\Pi_i$ encoding the topological information of the second Chern class $c_2(M_3)$. Choosing $\hat{D}_i$ as a basis of harmonic $(1,1)$-forms on $M_3$ one finds that
  \begin{equation}\label{def_pi}
   \Pi_i = \int_{M_3} c_2 \wedge \hat{D}_i \ .
 \end{equation}
With respect to an arbitrary choice of two-cycles, the numbers $\Pi_i$ can have both signs, and moreover they can vanish for some choices of moduli. For instance, the example of $K3$-fibered threefold $\mathbb{CP}^4_{1,1,1,6,9} [18]$ has
 \be
  \Pi_1 = 36, \qquad \Pi_2 = 0,   \qquad \Pi_3 = 0,   \qquad \Pi_4 = 0,   \qquad \Pi_5 = 102.
  \label{Pi} 
 \ee
We conclude that this class of corrections appears flexible in terms of signs and zeroes.

For the particular case of fibred Calabi-Yaus with two moduli, the resulting contributions to the scalar potential take the form 
 \begin{align}\label{effectivePotential}
   \delta V & = g_s^2\frac{W_0^4}{\cV^4}\left(- \mathcal{C}_1\frac{\cV
}{\tau_1} -\mathcal{C}_2\sqrt{\tau_1} \right)   = V_0\left(-\frac{\lambda^2\mathcal{C}_1 \tau_2^2}{\cV} -\frac{\mathcal{C}_2\cV}{\lambda\tau_2}   \right) \,,
  \end{align}
with $\mathcal C_i \sim \Pi_i$. One can consider the following possible interplays between such corrections (or a subset of them) and loop corrections:
 \begin{itemize}
\item
Inflation to the right  with 
\begin{equation}\label{effectivePotential1}
 \delta V = V_0\left(  -{\mathcal C_2\cV\over \tau_2}  +\frac{\cV^2}{g_s W_0^2}{(C_2^{KK})^2 \over \tau_2^2} \right) \,.
\end{equation}
As before, this leads to an $\alpha=2$ attractor.  Possible corrections proportional to e.g.~${\mathcal C_1}$ or $ {C_1^{KK}} $, are either absent for topological reasons or due to the choice of brane wrappings, or when present will modify the inflationary plateau similar to the discussion for loop corrections.
\item Inflation to the left with
\begin{equation}
\delta V = V_0 \left( -\frac{\mathcal C_1 \tau_2^2}{\cV} + \frac{(C_1^{KK})^2 \tau_2^4}{g_sW_0^2\cV^2} \right) \,.
\end{equation} 
In contrast to the general discussion of the previous section, this leads to an inflationary attractor with $\alpha = 1/2$. The reason is the absence of a linear term. In general,  with leading corrections of a higher $n$-th order, one obtains $\alpha = 2/n^2$. Again we are ignoring other corrections, which if present would modify the single-field nature.
\end{itemize}

\begin{itemize}

\item Finally, we can balance the higher superspace-derivative corrections against the string loop winding mode term~\cite{Cicoli:2016chb}. 
In that  case we get a potential
\begin{equation}
\delta V = V_0 \left(\frac{\mathcal C_1 \tau_2^2}{\cV} - \frac{C_{12}^W}{g_sW_0^2} \tau_2 \right) \,.
\end{equation}
\end{itemize}

\subsection{Scalar potential and dynamics from loop corrections: the generic case} 

In order to get an idea of where this happens for generic values, first assume that the minimum after inflation is determined by the first and last terms of the scalar potential \eqref{string_loops}, which fall off at infinity. The minimum is located at
 \begin{align}
   \tau_2^{3} = \frac{C_{12}^W \cV^2}{2\lambda^3(g_s C_{1}^{KK})^2} \,. \label{minimum}
 \end{align}
At this minimum, the second term with opposite behavior has a relative size of order
 \begin{align}
   \left( \frac{\lambda g_s^2 C_1^{KK} C_2^{KK}}{C_{12}^W} \right)^2 \,,
 \end{align}
which is assumed to be subdominant when the minimum is determined by the first two terms. However, it grows quadratically with decreasing with $\tau_2$. Therefore this ratio will become order one when $\tau_2$ has decreased with the square root of the inverse of the above ratio. It is there that the steepening of the potential becomes due to the $C_2^{KK}$ corrections dominant. In terms of the canonical inflaton, this corresponds to a steepening field range of
 \begin{align}
  \Delta \varphi_{\rm steep.} \simeq \sqrt{3} \log\left( \frac{C_{12}^W}{\lambda g_s^2 C_1^{KK} C_2^{KK}} \right) \,. \label{steepening}
 \end{align}
Every order of magnitude in the argument of the logarithm leads to a field displacement of $\sqrt{3} \log 10 \approx 4$. This clearly shows that one needs a non-trivial hierarchy in order to have a sufficiently long plateau to  sustain inflation.

An appealing manner to obtain such a range would be to have a very weak string coupling. However, this also leads to an exponentially large volume due to~\eqref{LVS}, which is incompatible with CMB observations. In particular, the COBE normalization of CMB temperature anisotropies requires the height of the scalar potential during inflation to be of the order $10^{-10}$. Note that this height scales as $\cV^{-10/3}$, given by the difference of the loop correction terms in \eqref{string_loops} at the minimum \eqref{minimum} and during inflation, where they vanish. Therefore natural values of the volume are around $10^3$ or $10^4$.

The above discussion also indicates what happens when the correction become important. The volume stabilization takes place at $\cV^{-3}$ and the inflationary dynamics just a factor $\cV^{-1/3}$ below this\footnote{Note that this crucially relies on the extended no-scale structure: with linear instead of quadratic corrections to $\delta V$, the inflationary dynamics would instead be a factor $\cV^{1/3}$ above the scale of volume stabilization.}. Due to the limited range for the volume, it is hard to separate these scales parametrically. One would therefore expect that at latest at the moment when the $C_2^{KK}$ corrections reach the volume modulus scalar potential scale, the volume stabilization also ceases to be effective and  the volume becomes a dynamical variable as well (see also \cite{Cicoli:2008gp}).

 \begin{figure}[h!]
\begin{center}
\includegraphics[scale=0.5]{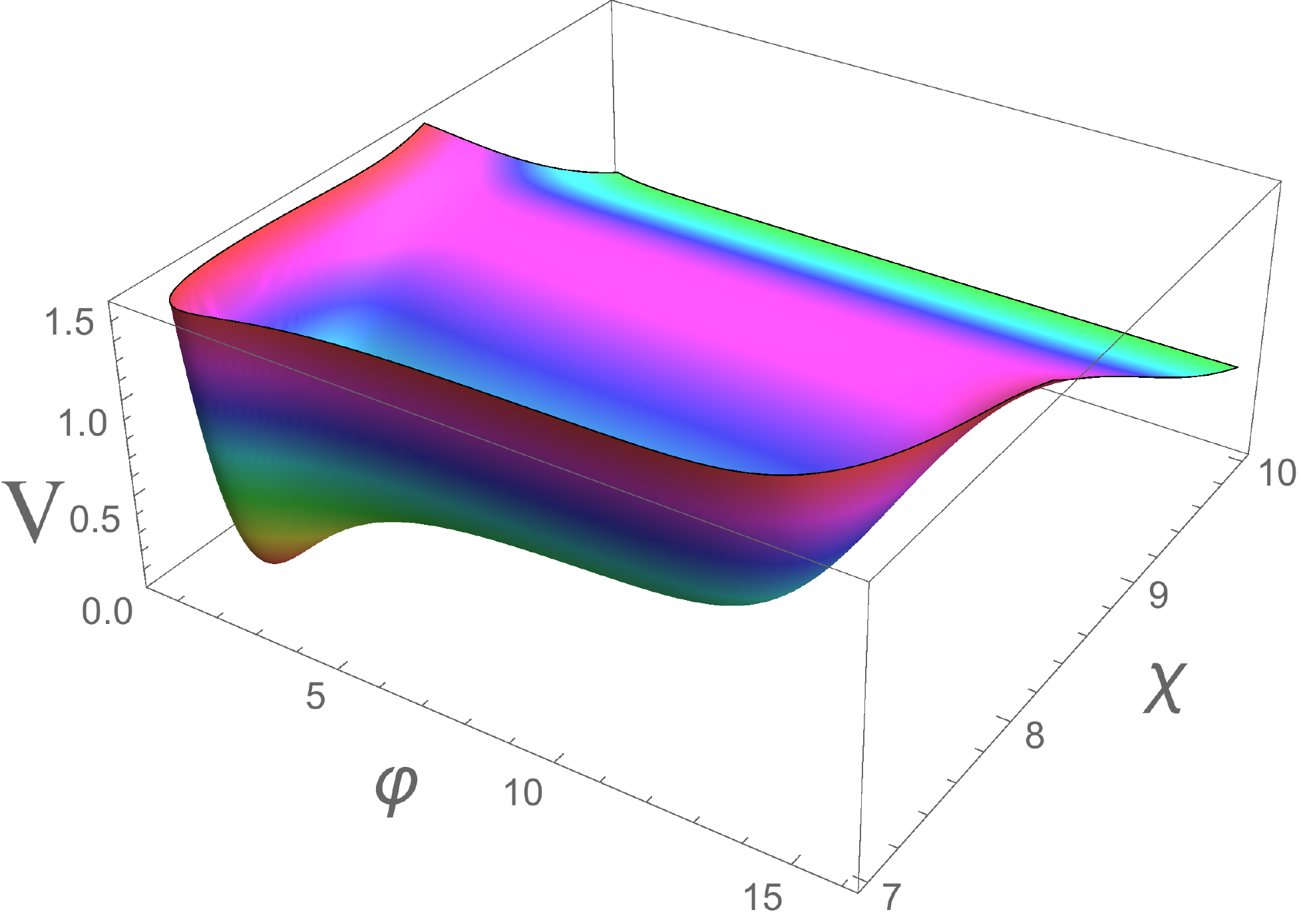}
\end{center}
\caption{\footnotesize The scalar potential $V$ of the Fibre inflation. For $\varphi>15$, the potential in this model begins to rise, whereas at large $\chi$ the potential falls down. }
\label{FOlarge}
\end{figure}

Therefore, beyond  
\begin{align}
  \Delta \varphi_{\rm 2-field} \simeq \sqrt{3} \log\left( \frac{C_{12}^W}{\lambda g_s^2 C_1^{KK} C_2^{KK}} \cV^{1/6}\right) =\Delta\varphi_{\rm steep.}+\frac{1}{2\sqrt{3}}\log\,\cV\simeq \Delta\varphi_{\rm steep.}+{\cal O}(1)\,, \label{breakdown2}
 \end{align}
one should not trust the picture with a scalar potential that bends upwards solely as a function of $\tau_1$; instead, the actual  dynamics is determined by a two-dimensional field space, see Fig.~\ref{FOlarge}. We note, however, in concordance with~\cite{Cicoli:2008gp} that already for field values $\varphi$ between the onset of steepening and the onset of 2-field dynamics, the slow-roll parameters increase so drastically due to steepening that slow-roll breaks down there. Hence, the whole slow-roll  region inside the scalar potential valley proceeds approximately with single-field dynamics.
Thus the process of inflation in the fibre inflation model occurs only in the certain range of the variables $\varphi$ and $\chi$, along the inflationary valley shown in Fig. \ref{FO} and Fig. \ref{FOlarge}. In particular, for sufficiently large values of $\chi$, the potential bends down, and the field $\chi$ starts to grow.

\subsection{The speculative case with fewer corrections: recovering the infinite $\alpha$-attractor plateau}

We do expect that at higher order in the $\alpha'$- and string loop $g_s$-expansion  singular terms  might  eventually arise in the scalar potential even if we were able to find setups where a part of the leading corrections is absent. This is because there is no manifest microscopic symmetry protecting the K\"ahler potential from K\"ahler moduli string loop corrections at any loop order. The infinite plateau $\varphi\to\infty$ corresponds to a 4-cycle $\tau_2\sim \exp(-\varphi/\sqrt 3)\to 0$ shrinking to zero whereas the volume of the K3-fibre $\tau_1\sim \exp( 2\varphi/\sqrt 3)\to \infty$ blows up.  No information is available  about  string corrections at all higher orders in this regime. We may speculate that such corrections will make the exponential plateau of fibre type to be of finite length, or we may speculate that under certain specific conditions, these unknown corrections will not affect the potential. 

Either way, if we speculate about particular setups where a part of  the leading order $\alpha'$- and $g_s$-corrections is absent, then for such setups  the plateau length can turn out to be much larger than inferred from the leading order $\alpha'$- and $g_s$-corrections.  We will now sketch the vanishing requirements of such infinite plateau setups, bearing in mind that we do not have explicit setups exhibiting the non-generic partial vanishing of the loop and/or higher superspace-derivative corrections.


\subsubsection{Loop corrections -- the idealized case: infinite plateau}

Let us look at the most simple case of a fibred CY with just 3 volume moduli at all, of which the first 2 comprise the fibred `LARGE' part of the volume $\lambda\sqrt\tau_1\tau_2$, and the 3rd must be a true del Pezzo blowup supporting the ED3 instanton necessary for LVS stabilization.

In this simplest case, the fibration structure ensures that the 4-cycles of the two K\"ahler moduli determining the product structure of the CY volume $\cV=\lambda \sqrt{\tau_1}\tau_2$ necessarily intersect with each other. Hence, if we tried to forbid the winding mode string loop corrections in $\tau_1$ and $\tau_2$ entirely, in this most simple case we might be able to do so by wrapping only one of the 4-cycles corresponding to $\tau_1$, $\tau_2$ with D7-branes.

So for the simplest class of fibred CYs, if we find a model where $C_{12}^W=0$ then we might expect that either $C_1^{KK}=0$ or $C_2^{KK}=0$, as far as the exchange of KK modes among the D7-branes wrapping $\tau_1$ and $\tau_2$ is concerned. Conversely, if we found a setup where $C_{12}^W\neq 0$ then this entails $C_1^{KK}=C_2^{KK}=0$, as now the $\tau_1$- and $\tau_2$-4-cycles intersect each other, forcing the KK-mode corrections from both cycles to vanish.  However, note that successful LVS stabilization requires even for the simplest fibred CY a 3rd pure del Pezzo blow-up modulus, which intersects only with itself, so it can carry an ED3 instanton. While this blow-up does not carry a D7-brane, it is parallel to the two divisors $\tau_1$ and $\tau_2$ and thus shares the same orthogonal two real dimensions as the two fibration 4-cycles. Hence, we would generically expect this to give rise to additional KK-mode corrections of the type $C_1^{KK}$ and $C_2^{KK}$.

 \begin{figure}[t!]
\begin{center}
\includegraphics[scale=0.5]{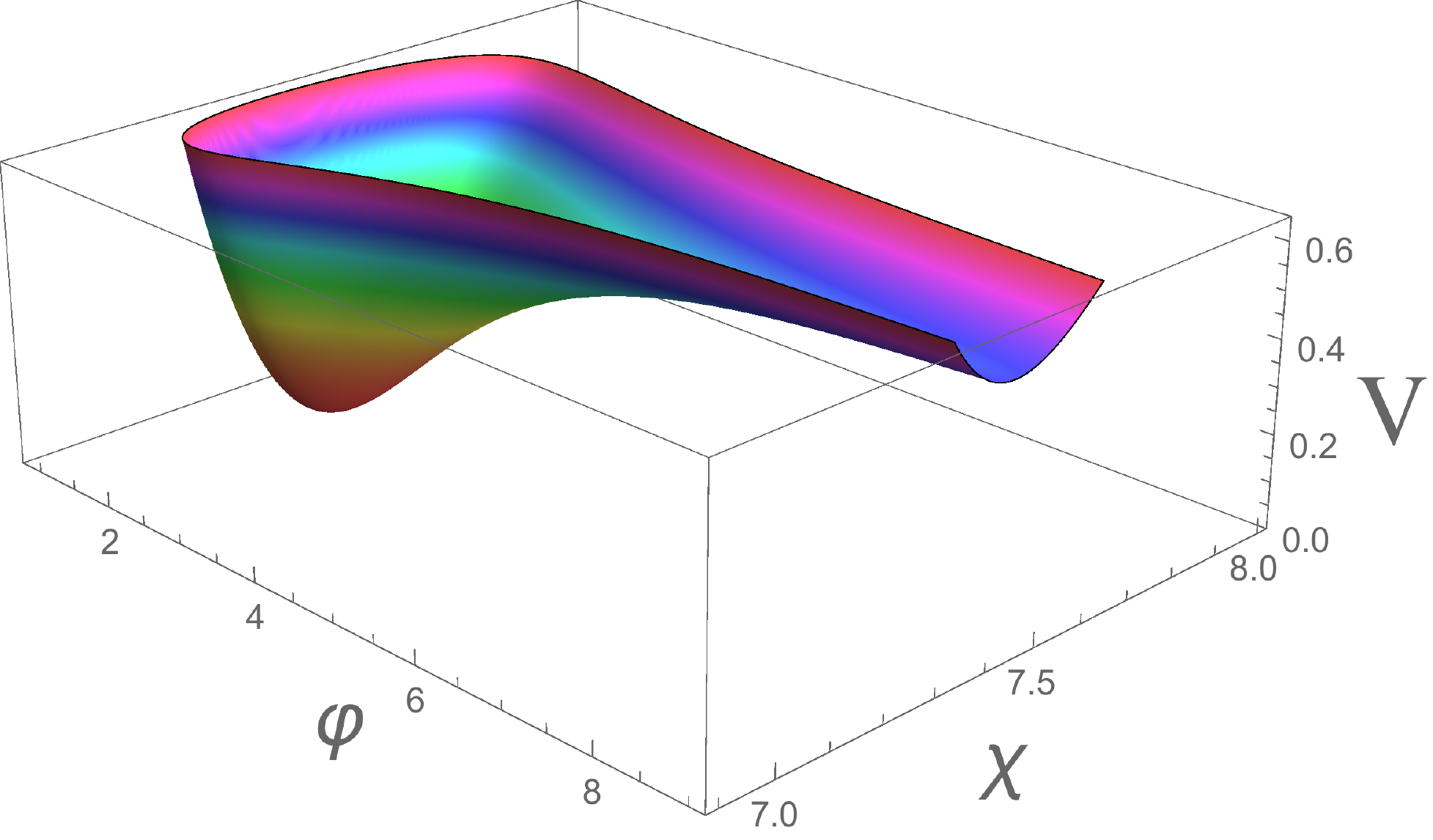}
\end{center}
\caption{\footnotesize The scalar potential $V$ of the fibre inflation with a particular set of parameters; $\chi$ is defined by $\chi=\log{\cal V}$.  }
\label{FO}
\end{figure}

Finally, we can discuss what happens in the absence of such corrections, at least in the observable window up to 60 e-folds. Ignoring $C_2^{KK}$ for the moment, upon including an uplift term leads to an inflationary potential with an infinite plateau at large $\varphi$, see Fig. \ref{FO}.  The leading deviation from this is given by the third term in \eqref{string_loops} and therefore of the form $\exp(- \varphi / \sqrt{3})$. If such setups can be found then they would lead to the robust inflationary predictions of $\alpha$-attractors \cite{Kallosh:2013yoa} with the specific value $\alpha = 2$, as discussed in \cite{Burgess:2016owb}.

\subsubsection{$F^4$ corrections -- the idealized case: infinite plateau}

We now see that once we grant the assumed particular minimal CY setups with or without the winding correction discussed in the previous subsection, we find no singular terms in both inflation to the left and to the right arising from the string loop corrections. The only corrections able to spoil the plateau with singular behaviour at small $\tau_2$ are the higher superspace-derivative corrections $\mathcal C_1$ or $\mathcal C_2$, respectively.


The vanishing of either $\mathcal C_1$ or $\mathcal C_2$ is a well defined model selection question. This is, because the higher superspace-derivative terms depend explicitly on the topological data of the second Chern class of the CY as well as the choice of K\"ahler cone. Hence, we see that if there existed fibred CYs conforming to the speculations of the previous subsection where in addition either $\Pi_1$ or $\Pi_2$ and consequently either $\mathcal C_1$ or $\mathcal C_2$ vanish, we can ensure the absence of rising singular terms which limit the plateau potential at the level of the leading $\alpha'$ and string loop corrections.

\subsection{General relation to $\alpha$-attractors}

Above we have seen that the general framework of fibre inflation shares many features with $\alpha$-attractors: in the absence of corrections that destroy the inflationary plateau, they are identical with specific values of $\alpha$, while corrections that grow in importance at large field values give rise to a multi-field generalization of $\alpha$-attractors. Let us outline the origin of this correspondence.

In the case of a product of hyperbolic manifolds, the general structure of $\alpha$-attractors can be defined by the K\"ahler and superpotential
\begin{align}
K=&-\log (T_1+\ol{T}_1)-2\log(T_2+\ol{T}_2)+S\ol S \,, \qquad
W= S f(T_1,T_2) \,.
\end{align}
Moreover, we assume that the volume stabilization condition $\tau_1\tau_2^2=\frac{1}{\lambda^2}\langle \cV_0^2\rangle$}  is already imposed by the previous stage of the theory.
At this point we study only inflation and will not specify the exit now, where $S$-independent terms in the superpotential and the question of taking $S$ nilpotent or just heavy become relevant.

The discussion now splits in two separate cases, depending on the functional dependence of $f$. First of all, one can assume that this function only depends on $T_2$, and is regular near $\text{Re} T_2 \rightarrow 0$. Restricting to vanishing axions\footnote{In examples one can check that the axions may need stabilization. In such case the extra geometric term in the \K\, potential, associated with the bisectional curvature, will do the job \cite{Carrasco:2015uma}. We can add the following type of terms $ S\bar S (T_i- \bar T_i)^2 F( \tau_j)$.}, this model has a kinetic and potential energy given by
\begin{align}
  -\frac32 \frac{\pd\tau_2\pd\tau_2}{\tau_2^2} - f^2 (\tau_2)  \,, \label{a=2}
\end{align}
where $f$ is a regular function at the pole around $\tau_2 \rightarrow 0$ (in fibre inflation this is achieved by a constant \K~potential due to volume stabilization).
The generic example of a regular function at  is  $1- c\tau_2 +\dots$, yielding an E-model  of the $\alpha =2$ attractor. All predictions are $c$-independent and follow from the leading term that breaks the non-compact symmetry (see \cite{Burgess:2014tja} for a discussion of the analogy to the compact symmetry of natural inflation). Examples of the above behaviour are provided by string loops \eqref{string_loops} as well as the right model with higher derivatives. These differ from the general structure \eqref{a=2} by having an expansion around $\tau_2 \rightarrow \infty$ rather than around zero; however, the above \K\ potential has an inversion symmetry $T_1 \rightarrow 1/T_1$ and $T_2 \rightarrow T_2$ which leaves the \K\ potential up to a volume-dependent shift, which we assume to be constant. Thefore the difference in expansion is immaterial for the predictions.
  
Alternatively, the function $f$ can give rise to a regular expansion in $T_1$ around the point $T_1 =0$. This yields the different behaviour
  \begin{align}
-\frac{3}{8}\frac{\pd\tau_1\pd\tau_1}{\tau_1^2}- f^2 (\tau_1).
\label{t1}\end{align}
Again, a generic regular function now at $\tau_1 \rightarrow 0$ is  $1- c\tau_1 +\dots$, and we get an E-model  of the $\alpha ={1\over 2}$ attractor, where $c$ again drops out. An example of this behaviour is the left model with higher derivatives. When phrased in terms of $T_1$, this exactly corresponds to a regular expansion, again in $1/T_1$ rather than $T_1$, which is not relevant due to the inversion symmetry.

The general case in which the function $f$ has a regular expansion in both $T_1$ and $T_2$ is fundamentally different. An expansion in both moduli is imcompatible with volume stabilization; when $T_1$ is small, $T_2$ blows up at fixed volume and vice versa. Therefore one has to include the dynamics of both moduli in such an expansion; the resulting inflationary scenario is multi-field in general.

In summary, the merger of two $\alpha$-attractors with $\alpha_i = (1/3,2/3)$ gives rise to a combined one with $\alpha=2$ or $\alpha = 1/2$, assuming volume stabilization. The choice between both $\alpha$'s is determined by the superpotential. More generally, the condition $\tau_1^{p_1} \tau_2^{p_2}$ fixed leads to a combined attractor with (more details can be found in appendix A)
 \begin{align}
   \alpha = \Big( \frac{p_2}{p_1} \Big)^2 \alpha_1 + \alpha_2 \,,
\end{align}
when expanding in $\tau_2$, or its inverse when expanding in $\tau_1$ (where we have assumed $\alpha_1 + \alpha_2 = 1$ in order to have a no-scale structure for the volume at lowest order). The values of $\alpha=2$ and $\alpha={1\over 2}$ in these models have a clear origin in the kinetic term structure of the $\mathbb{C}P^4_{[1,1,2,2,6]}  [12]$ model.

More generally, the dimensional reduction of type IIB string theory on a Calabi-Yau manifold dictates the tree level K\"ahler potential of the 2-cycle volume moduli to be given by a third-order homogeneous polynomial of the 2-cycle volumes $v^i$
\begin{equation}
K_{K}=-2\ln\,\cV\;\;,\;\;\cV=\frac16 \kappa_{ijk}v^iv^jv^k\;\;.
\end{equation}
The 4-cycle volumes $\tau_i$ are related to the 2-cycle volumes as
\begin{equation}
\tau_i=\frac{\partial\cV}{v^i}= \frac12 \kappa_{ijk}v^jv^k\;\;.
\end{equation}
Hence, for a fibred Calabi-Yau the dominant part of the volume will always take the form
\begin{equation}
\cV=\frac16 \kappa_{122}v^1(v^2)^2+\ldots\;\;{\rm or}\;\;\cV=\frac16 \kappa_{123}v^1v^2v^3+\ldots\;\;.
\end{equation}
Looking then at the relation between 2-cycle and 4-cycle volumes above, we see that the only possible values for $\tau_i$ powers in the fibration (product) part of the CY volume are $p_i=(1/2,1)$ implying $\alpha_i=(1/3,2/3)$. Hence, the limiting values $\alpha=(1/2,2)$ seem to be rather universal for the landscape of fibre inflation on CY compactifications of type IIB string theory. For the case of a general fibred Calabi-Yau with two volume moduli~\cite{Burgess:2016owb} we get $p_i=(1/2,1)$, hence $\alpha=(1/2,2)$ are the only unique possibilities (see Appendix~\ref{TwoModAlpha} for a detailed argument).

\section{$\overline{D3}$ induced geometric fibre model}


The effective supergravity model of fibre inflation can be given in the form suggested in \cite{Kallosh:2017wnt}.
The potential depends on the  \K\, function $\cG$ which, in general is of the form

\be
\cG \equiv K + \log W +\log \bar W\, , 
\qquad 
    \mathbf{ V} = e^ {\cal   G}  (\cG^{I \bar J }  {\cal   G} _I  {\cal   G} _{\bar J}  - 3 ).
    \ee
     In our case the index $I$ includes the directions   $ S$ and $T_i= ( T_1, T_2 )$.
     We take     
    \be
\cG (T_i, \bar T_i; S, \bar S) = \cG_0(T_i, \bar T_i) + S +\bar S + \cG_{S\bar S} (T_i, \bar T_i)S\bar S \,,
\label{cG0}\ee
and suggest  the following \K\, function for the fibre inflation:
\begin{align}
\cG=& \log |W_0|^2 -{1\over 2} \log \frac{(T_1+\overline{T}_1)^2}{4T_1\overline{T}_1} 
-\log \frac{(T_2+\overline{T}_2)^2}{4 T_2\overline{T}_2}  +
S+\overline S
+\cG _{S\bar S}(T_i, \bar T_i) S\overline S.
\end{align}
Here the $S$-field metric depends on a potential as follows
\be
\cG _{S\bar S}(T_i, \bar T_i)= \frac{m_{3/2}^2}{3 m_{3/2}^2+{\bf V}(T_1,\overline{T}_1,T_2,\overline{T}_2)}
\ee
where $m_{3/2}$ is the gravitino mass.
The potential consists of three terms
\be
{\bf V}(T_1,\overline{T}_1,T_2,\overline{T}_2)= \Lambda + V_{\rm stab} +  V_{\rm infl}.
\ee
The cosmological constant at the exit at the minimum of the potential is
\be
\Lambda = F_S^2- 3 m_{3/2}^2
\ee
where
\be
|F_S|^2\equiv |\cG_S|^2 \equiv e^{\cG_0} \cG_S \cG^{S\bar S} \cG_{\bar S}  \,.
\ee
We can now determine $V_{\rm stab}$ to lowest order by expanding out the LVS volume stabilization scalar potential in a quadratic neighborhood of the volume minimum $\langle\cV\rangle\equiv\cV_0$. If we denote the volume modulus mass as $M$, then
\begin{equation}
V_{\cV}=M^2 (\cV-\langle\cV\rangle)^2=M^2(\lambda \sqrt{\tau_1}\tau_2-\cV_0)^2\;\;.
\end{equation}
Hence, we will choose the form
\be\label{stab}
 V_{\rm stab} = M^2\left(\frac{\lambda}{8}(T_1+\bar T_1) (T_2+\bar T_2)^2-\cV_0^2\right)^2
\ee
for the volume stabilization potential, since this clearly reproduces $V_\cV$ in it own quadratic neighborhood.
The mass parameter $M$ is assumed to be significantly larger than the scale of a cosmological term $V_{\rm infl}$, and from now on we put $\lambda/8=1$ for simplicity. This would correspond to a spirit of the original fibre inflation model with a strong stabilization of the large volume of compactification, such that stringy corrections responsible for a cosmological evolution do not affect stabilization of the total volume.

We can now incorporate the scalar potential for $\tau_1$ and $\tau_2$ using a similar comparison with the actual fibre models we did above for the overall volume stabilization. In a quadratic neighborhood of the full fibre inflation scalar potential the scalar potential for $\tau_1$ and $\tau_2$ will read
\begin{equation}
V_{\tau_1}=m^2 \left(\langle\tau_1\rangle-\tau_1\right)^2
\end{equation}
and
\begin{equation}
V_{\tau_2}=m^2 \left(\langle\tau_2\rangle-\tau_2\right)^2\;\;,
\end{equation}
respectively. If we now, for simplicity, rescale their minima $\langle\tau_i\rangle$ to unity, then we can clearly take the cosmological part of the potential in the simplest interesting cases with $\alpha=2$ and $\alpha=1/2$, respectively, as follows
\be
 V_{\rm infl}^{\alpha=2}= m^2\left(1-\frac12(T_2+\overline{T}_2)\right)^2,
\ee

\be
 V_{\rm infl}^{\alpha=1/2}= m^2\left(1-\frac12(T_1+\overline{T}_1)\right)^2.
\ee

We discuss the stability of non-inflaton directions during inflation. In the following discussion, we will use $V_{\rm infl}^{\alpha=2}$ as the inflaton potential. Because of the stabilizing term in the scalar potential, we introduce the following new basis,
\begin{align}
\varphi=-\frac{1}{\sqrt3}(\sqrt{2}u_1-u_2), \quad \chi=\frac{1}{\sqrt3}(u_1+\sqrt2 u_2),\quad \theta=\frac{1}{\sqrt3}(\sqrt{2}a_1-a_2),\quad \psi=\frac{1}{\sqrt3}(a_1+\sqrt2 a_2),
\end{align}
where $u_i$ and $a_i$ are defined by $T_i=e^{\sqrt{\frac{2}{3\alpha_i}}u_i}(1+{\rm i}\sqrt{\frac{2}{3\alpha_i}}a_i)$, and $3\alpha_i=i$ for $i=1,2$. Both $(\varphi,\chi,\theta,\psi)$ and $(u_i,a_i)$ are canonical on inflationary trajectory $a_i=0(=\theta=\psi)$. In the limit $m\to0$, we find $\varphi$ is a flat direction, and the minimum is given by $\chi=\chi_0=\frac{1}{\sqrt 6}\log \frac{\cV_0^2}{8}$ and $\theta=\psi=0$. At $\chi=\chi_0$, the inflaton potential becomes the E-model $\alpha$-attractor potential
\begin{align}
V_{\rm infl}|_{\chi=\chi_0}=V_{\rm eff}=m^2\left(1-\frac{\cV_0^{2/3}}{2}e^{-\frac{2\varphi}{\sqrt3}}\right)^2.
\end{align}
\begin{figure}[h!]
\begin{center}
\includegraphics[scale=0.5]{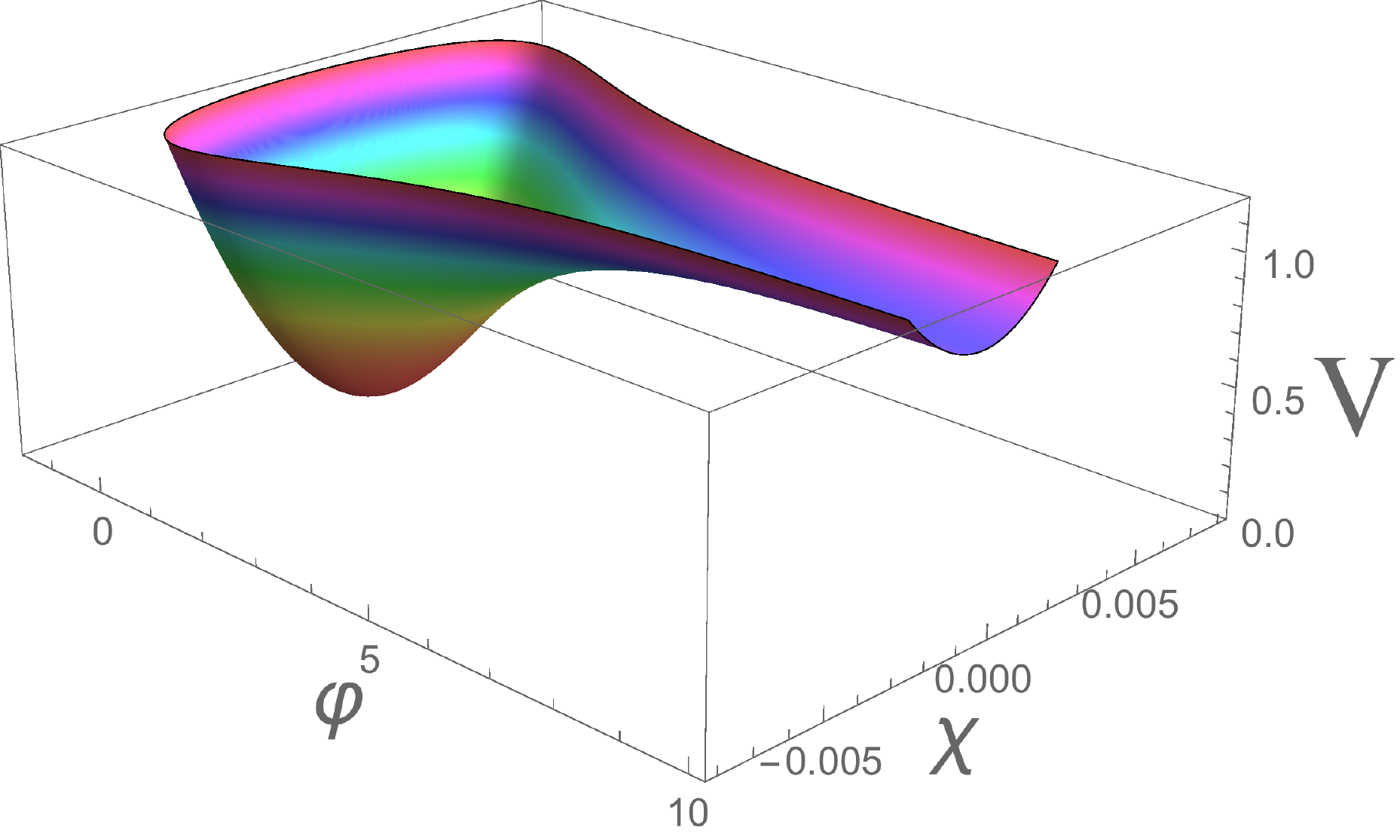}
\end{center}
\caption{\footnotesize E-model version of  the fibre inflation potential. }
\label{2E}
\end{figure}

Due to the inflationary potential, however, the minimum of $\chi$ is slightly shifted from $\chi=\chi_0$. The scalar potential at $\theta=\psi=0$ is shown in Fig.~\ref{2E}. The deviation gives extra contribution to the scalar potential as
\begin{align}
\delta V=\frac{\epsilon^2}{36\cV_0^{8/3}}V_{\rm eff},
\end{align}
at the leading order of the $\epsilon$ expansion, where $\epsilon=\frac{m}{M}$. This contribution is negligible for $\epsilon\ll1$, and we will neglect it in the following discussion.

The mass of the axionic directions $\theta$ and $\psi$ are given by
\begin{align}
m_{\theta}^2&=m_{\psi}^2=4W_0^2+2V_{\rm eff},
\end{align}
which are positive definite during inflation. The heavy modulus $\chi$ has the mass
\begin{align}
m_{\chi}^2=12M^2\cV_0^4.
\end{align}
Note that all the masses are the leading part of the $\epsilon$-expansion. The minimum of the potential is given by $\varphi=\frac{\sqrt{3}}{2}\log\frac{\cV_0^{2/3}}{2}$, and the masses are given by
\begin{align}
m_\varphi^2=2m^2,\quad m_{\chi}^2=12M^2\cV_0^4,\quad m_{\theta}^2&=m_{\psi}^2=4W_0^2.
\end{align}
 Thus, we can conclude that this system is stable during and after inflation.
 
Note the similarity of the inflaton potential in the $\alpha$-attractor model considered in this section and shown in Fig. \ref{2E} to the potential of   fibre inflation shown in Fig. \ref{FO}. This is in accord with our expectations that our supergravity model correctly captures essential features of fibre inflation in the vicinity of the inflationary trajectory.

\section{Discussion}

The increasingly precise data from the cosmic microwave background (CMB) during recent years provide very strong observational support for an early phase of cosmological inflation. At the same time the data starts to put relevant upper bounds on the tensor-to-scalar ratio $r<0.07\,(95\,\%)$. 

Given this situation, it is interesting to study bottom-up inflation models which are both simple and at the same time cover a wide class of potentials, while providing suppressed levels of tensor modes in the regime $10^{-3}< r < 10^{-2}$ and maintaining a good fit to the observed value of spectral tilt $n_s\simeq 0.97$. Since these levels of  $r$ imply a very high scale of inflation, we should at the same time aim for bottom-up inflation models which have a possible UV completion in models of string inflation. 

$\alpha$-attractors~\cite{ Kallosh:2013yoa} are a very general class of such inflation models constructed bottom-up in 4D ${\cal N}=1$ supergravity.  They produce exponential plateau potentials controlled by a single parameter $\alpha$ labeling the residue of a second-order pole of the kinetic term of the inflaton. Due to the presence of this pole, $\alpha$-attractor models are `pole inflation' models~\cite{Galante:2014ifa,Broy:2015qna} which shift the question of quantum corrections affecting the inflationary dynamics  from the  scalar potential to the kinetic function. As long as the kinetic function is dominated by a second-order pole, an arbitrary analytic scalar potential will flatten out to yield an exponential plateau inflation with a universal prediction $n_s = 1 - 2 / N$ and $r = 12\alpha / N^2$ at $N$ e-folds before the end of inflation. 

However, despite their simplicity and generality $\alpha$-attractors so far had no clear link to a UV completion in string theory.
One of the main problems has been, that those string moduli fields, which acquire a second order pole in their kinetic function, often appear with pole at the same position in the scalar potential due to Weyl rescaling of the sources of the moduli potential into 4D Einstein frame.  In such cases, pole inflation  looses its flat plateau; for certain combinations of the orders of the poles in the kinetic function and the scalar potential this can even render inflation impossible. 

Yet, there are models of inflation in type IIB string theory compactified on Calabi-Yau manifolds, which combine polynomial potentials for certain volume moduli with a second-order pole in the kinetic term of these moduli. These `fibre inflation models'~\cite{Cicoli:2008gp,Burgess:2016owb} produce an exponentially flat plateau with a field range of ${\cal O}(5\ldots10\,M_{\rm P})$ in the extant semi-explicit toy model constructions.  

In this work we demonstrated that the low-energy effective description of the string models of `fibre inflation' are a class of $\alpha$-attractors. Moreover, we showed how the recently developed method of geometrizing $\alpha$-attractors using nilpotent superfields in supergravity~\cite{Kallosh:2017ced,Kallosh:2017wnt} allows us to write a simple and explicit 4D supergravity realization of the core dynamics of moduli stabilization and inflation in fibre inflation. 

Our supergravity realization of fibre inflation as an $\alpha$-attractor makes it clear, how a stringy realization of pole inflation can work: Namely, the LVS scenario inspired volume stabilization on a fibered Calabi-Yau manifold stabilizes the \emph{whole} Calabi-Yau volume, which is a \emph{product} of two volume moduli. This product-type of constraint from moduli stabilization allows for second-order poles in the kinetic functions of the individual moduli while keeping \emph{the K\"ahler potential constant along the inflaton direction given by one of the two volume moduli}. This way, fibre inflation is a stringy $\alpha$-attractor model which avoids the pole in the scalar potential from Weyl rescaling proportional to $e^K$.

As long as the total volume remains stabilized, each of the two volume moduli comprises an $\alpha$-attractor direction. Applying the fusion rules for $\alpha$-attractors with several fields studied in~\cite{Kallosh:2017ced,Kallosh:2017wnt}, and applying the general structure of the Calabi-Yau volume expressed in 4-cycle moduli $\tau_i$, we find that fibre inflation realizes $\alpha$-attractors with only two discrete values $\alpha= 1/2$ or $\alpha= 2$. 

This is valid, as long as the inflationary dynamics is effectively single-field keeping the total Calabi-Yau volume stabilized. We analyze the effect which the presence of higher-order corrections such as those conjectured to arise from string loops has on the exponential plateau. If they are present, then they lead to steepening of the potential after some finitely long exponential plateau. This steepening region very quickly increases the inflation potential to scale of the total volume stabilization. Beyond this point the dynamics becomes a 2-field model involving one of the two chosen $\alpha$-attractor directions and the volume modulus which becomes dynamical. We leave a study of this 2-field dynamics and its effect on the effective range of values of $\alpha$ as a very interesting subject for the future. 

 Finally, we also note that in some cases the dominant higher-order corrections may be absent. This may lead to the existence of very long inflationary flat directions.

\vspace*{1cm}

 \noindent{\bf {Acknowledgments:}} We are grateful to  C. Burgess, M. Cicoli, S. Parameswaran, F. Quevedo, and I. Zavala for stimulating discussions.  The work  of  RK, AL and YY is supported by SITP and by the US National Science Foundation grant PHY-1316699. The work of AW is supported by the ERC Consolidator Grant STRINGFLATION under the HORIZON 2020 grant agreement no. 647995. The work of AL is also supported by the Templeton foundation grant ``Inflation, the Multiverse, and Holography''. AW and DR are grateful to SITP for the hospitality when this work was initiated. All authors are grateful to the Lorentz center in Leiden, where the final part of this work was performed during the Lorentz workshop `Theoretical Approaches to Cosmic Acceleration'.

\appendix

\section{Fusion rule of $\alpha$}

In this section we will generalize the analysis of possible $\alpha$'s for generic two-moduli $\alpha$-attractors; see Appendix~\ref{TwoModAlpha} for the restrictions in actual Calabi-Yau compactifications.

Suppose we have two chiral superfields $T_1$ and $T_2$ with the K\"ahler potential given by
\begin{align}
-3\alpha_1\log(T_1+\ol{T}_1)-3\alpha_2\log (T_2+\ol{T}_2).
\end{align}
In this case, the `volume' $\tau_1^{3\alpha_1}\tau_2^{3\alpha_2}$ is invariant under the dilatation transformation
\begin{align}
&T_1\to \lambda^{\alpha_2} T_1, \qquad T_2\to \lambda^{-\alpha_1}T_2.
\end{align}
In terms of the canonical real variables $(u_i, a_i)$, defined as
 \begin{align}
   T_i=e^{-\sqrt{\frac{2}{3\alpha_i}}u_i}(1+{\rm i}\sqrt{\frac{2}{3\alpha_i}}a_i) \,,
 \end{align}
it is useful to perform the following field basis change:
\begin{align}
\chi=\frac{1}{\sqrt{\alpha_1+\alpha_2}}(\sqrt{\alpha_1}u_1+\sqrt{\alpha_2}u_2)  \,, \qquad \phi=\frac{1}{\sqrt{\alpha_1+\alpha_2}}(-\sqrt{\alpha_2}u_1+\sqrt{\alpha_1}u_2),
\end{align}
where $\chi$ is the invariant field under the dilatation, corresponding to the ``volume'' and $\phi$ is the orthogonal direction corresponding to the ``fibre''. In terms of the latter, which will provide the inflaton direction, the scalar potential reads
\begin{align}
V=V(T_1,T_2)=V\left(e^{-\sqrt{\frac{2}{3\tilde{\alpha}_1}}\phi},e^{\sqrt{\frac{2}{3\tilde{\alpha}_2}}\phi}\right),
\end{align}
 where
\begin{align}
&\tilde{\alpha}_1=\frac{\alpha_1}{\alpha_2}\left(\alpha_1+\alpha_2\right)\,, \qquad \tilde{\alpha}_2=\frac{\alpha_2}{\alpha_1}\left(\alpha_1+\alpha_2\right).
\end{align}
If the potential is effectively given by a polynomial of $T_i$, the model effectively becomes an attractor with $\alpha=\tilde{\alpha}_i$. For example, $\alpha_1=1/3$, $\alpha_2=2/3$ yield $\tilde{\alpha}_1=1/2$ and $\tilde{\alpha}_2=2$, which corresponds to the fibre inflation setups. Moreover, note that when the volume modulus has a no-scale structure, implying $\alpha_1 + \alpha_2 = 1$, then both resulting values of $\alpha$ are always inversely related.

Finally, one can consider a further generalization, which we will discuss in a simplified toy model without SUSY. We consider the Lagrangian
\begin{align}
-3\alpha_1\frac{\pd\tau_1\pd\tau_1}{4\tau_1^2}-3\alpha_2\frac{\pd\tau_2\pd\tau_2}{4\tau_2^2}-V_{\text{inf}}(\tau_1,\tau_2)-V_{\text{fix}}(\tau_1,\tau_2).
\end{align} 
$V_{\text{fix}}$ gives a constraint on $\tau_1$ and $\tau_2$, which we will assume to take the form $\tau_1=c\tau_2^p$ where $p$ and $c$ are constants.  In the fibre inflation case, this corresponds to the volume stabilization $\tau_1=\cV_0^2\tau_2^{-2}$. In terms of canonical variables $u_i$ defined by $\tau_i=e^{-\sqrt{\frac{2}{3\alpha_i}}u_i}$, the constraint reads
\begin{align}
u_1-\sqrt{\frac{\alpha_1}{\alpha_2}}pu_2=\text{const} \,.
\end{align}
Again we can decompose $u_i$ into the fixed mode $\chi$ and the flat mode $\phi$ as
\begin{align}
&\chi=\frac{\sqrt{\alpha_2}}{\sqrt{\alpha_2+p^2\alpha_1}}\left(u_1-\sqrt{\frac{\alpha_1}{\alpha_2}}pu_2\right) \,, \qquad 
\phi=\frac{\sqrt{\alpha_2}}{\sqrt{\alpha_2+p^2\alpha_1}}\left(\sqrt{\frac{\alpha_1}{\alpha_2}}pu_1+u_2\right).
\end{align}
The inverse relations read
\begin{align}
u_1=\frac{1}{\sqrt{\alpha_2+\alpha_1p^2}}\left(\sqrt{\alpha_1}p\phi+\sqrt{\alpha_2}\chi\right) \,,
\quad u_2=\frac{1}{\sqrt{\alpha_2+\alpha_1p^2}}\left(\sqrt{\alpha_2}\phi-\sqrt{\alpha_1}p\chi\right) \,.
\end{align}
Since $\chi$ is already fixed by $V_{\text{fix}}$, $\phi$ is the inflaton mode and its potential becomes
\begin{align}
V_{\text{inf}}(\tau_1,\tau_2)=V\left(e^{-\sqrt{\frac{2}{3\tilde{\alpha}_1}}\phi},e^{-\sqrt{\frac{2}{3\tilde{\alpha}_2}}\phi}\right) \,, \quad
\tilde{\alpha}_1=&\alpha_1+p^{-2}\alpha_2 \,, \quad 
\tilde{\alpha}_2=\alpha_2+\alpha_1p^2 \,.
\end{align}
Note that $p=1$ was the focus of \cite{Kallosh:2017ced} while in this paper we have investigated $p=-2$.

\section{Volume moduli dependence of fibre inflation with two moduli}\label{TwoModAlpha}

The classical Calabi-Yau volume is a cubic polynomical in the 2-cycle volumes
\begin{equation}
\cV = \frac16 \kappa_{ijk}v^iv^jv^k
\end{equation}
with $\kappa_{ijk}$ being the intersection numbers, topological numbers determined by the given CY.
For the case of a CY with two moduli the volume takes the most general form
\begin{equation}
\cV = \frac16 \kappa_{112}(v^1)^2v^2+\frac16 \kappa_{122}v^1(v^2)^2+\frac16 \kappa_{222}(v^2)^3
\end{equation}
where we have absorbed the $(v^1)^3$ piece into a shift of $v^2\to v^2+c v^1$, allowing us to set $\kappa_{111}=0$. The purest form of fibration clearly would have only $\kappa_{122}$ non-vanishing.

The relation between the 4-cycles $\tau_i$ and 2-cycles $v^i$ is given by
\begin{equation}
\tau_i = \frac{\partial\cV}{\partial v^i}=\frac 12 \kappa_{ijk}v^jv^k\quad.
\end{equation}
For our two-moduli case this system of coupled quadratic equations in the $v^i$ reads
\begin{align}
\tau_1= \frac13 \kappa_{112} v^1 v^2+\frac16 \kappa_{122} (v^2)^2 \,, \qquad
\tau_2= \frac16 \kappa_{112} (v^1)^2+\frac13 \kappa_{122} v^1 v^2 +\frac12 \kappa_{222} (v^2)^2 \,.
\end{align}
We wish to express $\cV$ in terms of the 4-cycle volumes, so we need to invert this system, solving for the $v^i$ as functions of the $\tau_i$. This can be done analytically, but the expressions are lengthy. Our interest is in the behavior solutions in the two fibre inflation asymptotic regimes $\tau_1\to\infty\;,\;\tau_2\to 0$ and $\tau_1\to 0\;,\;\tau_2\to \infty$ keeping $\cV$ constant, while we do not assume a particular relation between $\tau_1$ and $\tau_2$ at this point. We can then asymptotically expand the solutions $v^i(\tau_j)$ to the quadratic equations in $\tau_1$ and $\tau_2$ in these two regimes, and expand the solutions in $\kappa_{112}$ and $\kappa_{222}$ treated as perturbations to the pure fibration case where only $\kappa_{122}\neq 0$. We do this ony to analyze the scaling structure of the solutions, while it is clear that in reality intersection numbers given by topological data can never be an arbitrarily small continuous quantity.

In the regime $\tau_1\to 0\;,\;\tau_2\to \infty$ the solutions to the quadratic system are
\begin{align}
v^1 & = \frac{\kappa_{112}}{4\kappa_{122}}\sqrt{\frac{3}{2\kappa_{122}}} \frac{\tau_2^2}{\tau_1^{3/2}}+\sqrt{\frac{3}{2\kappa_{122}}} \frac{\tau_2}{\sqrt{\tau_1}} + \ldots \,, \notag \\ 
v^2 & =  \frac{\kappa_{112}}{\kappa_{122}}\sqrt{\frac{3}{2\kappa_{122}}} \frac{\tau_2}{\sqrt{\tau_1}}+\sqrt{\frac{6}{\kappa_{122}}}\sqrt{\tau_1} + \ldots \,,
\end{align}
where the dots represent higher-order terms. We see that $\kappa_{112}$ entails, that in the asymptotic limit $\tau_1\to 0$ both $v^1$ and $v^2$ would blow up and thus violate the constraint $\cV=const.$ From this perspective alone, a viable fibre inflation behavior would require $\kappa_{112}$ to be of very small magnitude. 

Fortunately, the analysis of~\cite{Burgess:2016owb} already argues that the most general two-moduli Calabi-Yau with a fibration structure has $\kappa_{112}=0$. From this we conclude that for  the general case of fibred CY we have
\begin{align}
v^1 = \sqrt{\frac{3}{2\kappa_{122}}} \frac{\tau_2}{\sqrt{\tau_1}}+\ldots \,, \qquad
v^2 = \sqrt{\frac{6}{\kappa_{122}}}\sqrt{\tau_1}+\dots
\end{align}
in the limit $\tau_1\to 0$, $\tau_2\to\infty$. This immediately implies $\cV\sim \sqrt{\tau_1}\tau_2$ in this limit which in turn dictates $\alpha=1/2$ for $\tau_1\to 0$, $\tau_2\to\infty$.

In the opposite regime $\tau_1\to\infty\;,\;\tau_2\to 0$ the solutions to the quadratic system are
\begin{align}
v^1 = \sqrt{\frac{3}{2\kappa_{122}}} \frac{\tau_2}{\sqrt{\tau_1}}-3\sqrt{\frac{3}{2\kappa_{122}}}\kappa_{222}\sqrt{\tau_1}+\ldots \,, \qquad
v^2 =  \sqrt{\frac{6}{\kappa_{122}}}\sqrt{\tau_1}+\dots \,,
\end{align}
where again the dots represent the higher-order terms. Here we see that for growing $\tau_1$ a non-vanishing $\kappa_{222}$ implies that there is maximum value of $\tau_1$ beyond which $v^1 < 0$. This violates the K\"ahler cone conditions for the given CY, which at minimum dictate that $v^i > 0$ for all 2-cycle volumes $v^i$ \emph{simultaneously}, see e.g.~\cite{Denef:2008wq}. Therefore, a fibred CY must have $\kappa_{222}=0$ in order to be `\emph{K\"ahler cone viable}' for fibre inflation. However, again, in that case our solutions become
\begin{align}
v^1 = \sqrt{\frac{3}{2\kappa_{122}}} \frac{\tau_2}{\sqrt{\tau_1}}+\ldots \,, \qquad
v^2 =  \sqrt{\frac{6}{\kappa_{122}}}\sqrt{\tau_1}+\dots \,,
\end{align}
and we have $\cV\sim \sqrt{\tau_1}\tau_2$ asymptotically. For the current regime $\tau_1\to\infty\;,\;\tau_2\to 0$ this implies $\alpha=2$.

Taken together, these two arguments imply that a two-volume-moduli CY which is `\emph{K\"ahler cone viable}' for fibre inflation, will always have a volume expression which asymptotically scales as $\cV\sim \sqrt{\tau_1}\tau_2$. Any such two-moduli CY with is `\emph{K\"ahler cone viable}' for fibre inflation should have  $\kappa_{222}=0$, which thus forms a condition for the  search for explicit CY examples of fibre inflation. Hence, for fibre inflation with two volume moduli there is a unique prediction of two discrete possibilities for $\alpha$, namely $\alpha=(1/2\,,\,2)$.

\section{T-model}

The string theory fibre inflation model discussed above is formulated in half-plane variables, suitable for the description of E-model $\alpha$-attractors. However, in terms of our effective supergravity approach, one can easily generalize this model, formulate it in disk variables, and find its version with the T-model potential:
\begin{align}
&{\cal G}=\log W_0^2- \ \log\frac{1-|Z_1|^2}{|1-Z_1^2|}-  2\log\frac{1-|Z_2|^2}{|1-Z_2^2|}+S+\overline{S}+g_{S\bar S}S\overline{S}\ ,\\
&g^{S\bar S}=\frac{1}{W_0^2}\left(|F_S|^2+{\bf V} \right).
\end{align}

As an example, one may consider the  scalar potential   
\be
{\bf V}(Z_{1},\overline{Z}_1,Z_{2},\overline{Z}_2)= \Lambda+ \frac{m^2}{2} |Z_2|^2 + V_{\rm stab}\ ,
\ee
where we will use the same stabilization potential $V_{\rm stab}$ as in the E-model \rf{stab}, but we represent it in terms of the disk variables, $T_{i} \to \frac{1+Z_{i}}{1-Z_{i}}$:
\be
V_{\rm stab}=  8M^{2}\, \frac{(1-|Z_1|^2)(1-|Z_2|^2)^{2}}{|1-Z_1|^2 |1- {Z}_2|^4} \ .
\ee 
Here $Z_{1} = \tanh{{\phi_{1} + i \theta_{1}}\over\sqrt{2}}$, $Z_{2} = \tanh{{\phi_{2} + i \theta_{2}}\over 2}$.

\begin{figure}[t!]
\begin{center}
\includegraphics[scale=0.5]{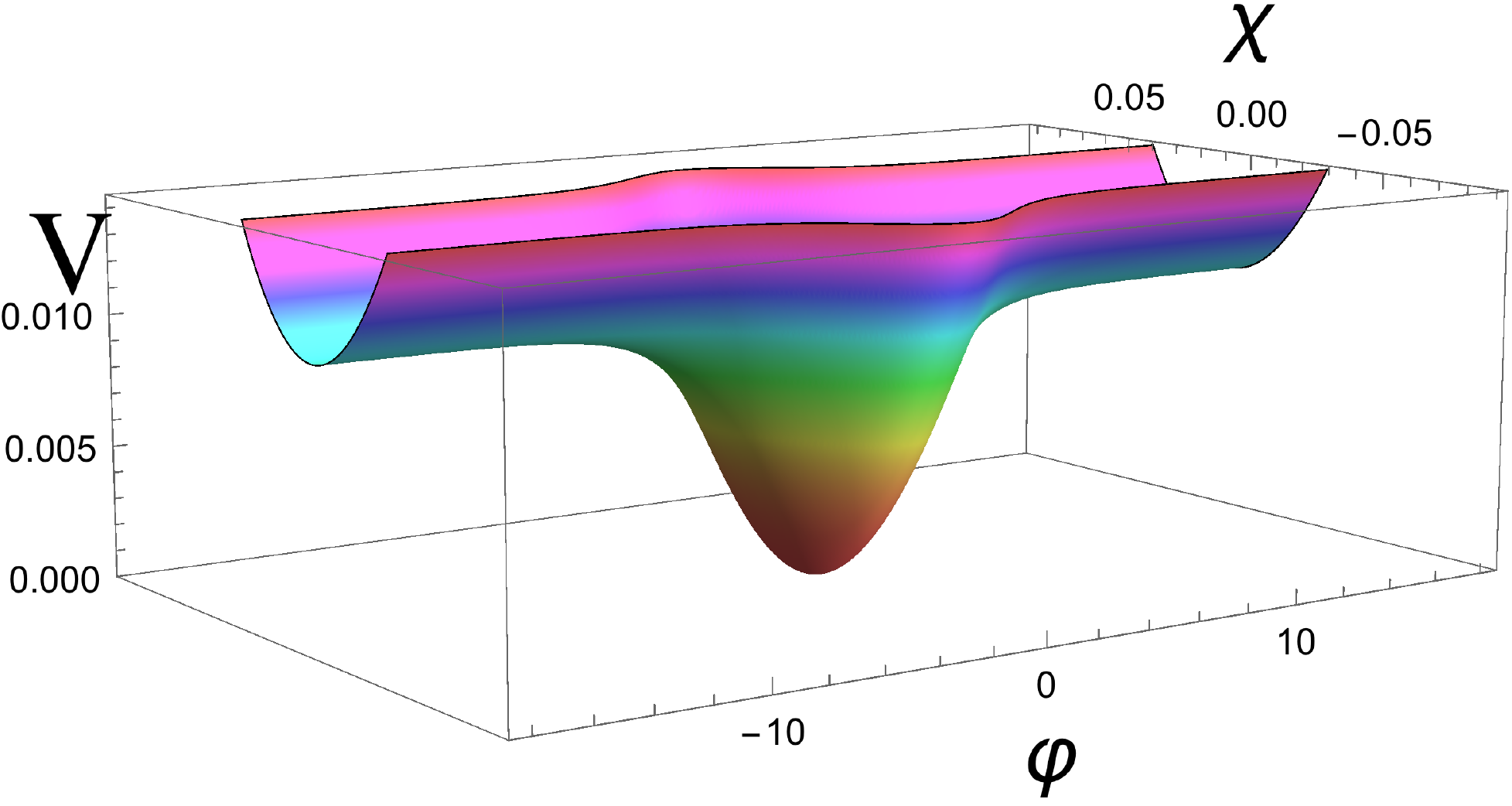}
\end{center}
\caption{\footnotesize Inflaton potential of the T-model version of the fibre inflation potential for a particular case $M =m= 0.1$, $\cV_0 = 2\sqrt 2$. }
\label{2T}
\end{figure}

It is convenient to express these fields in terms of their combination suitable for describing  inflation in this model:
\be
\phi_{1} =  \frac{\sqrt{2} \vp + \chi }{\sqrt 3},\quad \phi_{2} =\frac{\sqrt{2} \chi -\vp}{\sqrt{3}} ,\quad \theta_{1} =  \frac{\vartheta + \sqrt{2}\psi }{\sqrt 3},\quad \theta_{2} =\frac{\sqrt 2\vartheta -\psi }{\sqrt{3}} .
\ee
The total potential in terms of these fields, for $\Lambda = 0$, is 
\begin{equation}
\begin{split}
V=\sec ^2\frac{\sqrt{2} \vartheta -\psi }{\sqrt{3}} \sec \sqrt{\frac{2}{3}}
\left(\sqrt{2} \psi +\vartheta \right)
   \left[\frac{m^2 \left(\sinh ^2\frac{\varphi
   -\sqrt{2} \chi }{\sqrt{3}}+\sin ^2\frac{\sqrt{2} \vartheta -\psi
   }{\sqrt{3}}\right)}{\left(\cosh \frac{\varphi -\sqrt{2} \chi }{\sqrt{3}}+\cos
   \frac{\sqrt{2} \vartheta -\psi }{\sqrt{3}}\right)^2} 
   \right.
\\ 
\left.
+M^2 \left({\cV_0}^2 - 2
   e^{\sqrt{6} \chi } \cos \frac{\sqrt{2} \vartheta -4 \psi }{\sqrt{3}}-4 e^{\sqrt{6} \chi
   } \cos  \frac{2 \psi +\sqrt{2} \vartheta }{\sqrt{3}}-2 e^{\sqrt{6} \chi } \cos
   \sqrt{6} \vartheta \right)^2\right].
   \end{split}
\end{equation}

We will concentrate now on the inflaton potential with $\vartheta = \psi = 0$, which is given by a much simpler equation:
\be\label{real}
V=m^2 \tanh ^2\left(\frac{\varphi -\sqrt{2} \chi }{2 \sqrt{3}}\right)+M^2 \left(\cV_0^2-8
   e^{\sqrt{6} \chi }\right)^2 \ .
\ee
Let us now explore the general properties of this potential.  First of all, in the limit $M\cV_0^{2} \gg m$, the field $\chi$ tends to fall down to $\chi =  {2\over \sqrt{6}}\log{\cV_0 \over 8}$.  Then the potential of the field $\vp$ is given by the first term of \eqref{real}, which describes T-model $\alpha$-attractor shown in Fig. \ref{2T}.  An evaluation of the kinetic term of the field $\vp$ implies that it is an $\alpha$-attractor with $\alpha = 2$.

Now let us look at the same potential in the limit $\chi \to -\infty$, which brings us far away from the inflationary valley we just discussed. In this limit the potential becomes 
\be\label{reallimit}
V=m^2 \tanh ^2\left(\frac{\varphi -\sqrt{2} \chi }{2 \sqrt{3}}\right)+M^2 \cV_0^4 \,.
\ee
The minimal value of this potential on the  upper plateau is $M^2  \cV_0^4$. It is achieved for $\varphi =\sqrt{2} \chi$, 
which corresponds to $\phi_{2} = 0$. This direction is shown as a shallow blue valley on top of an infinite dS plateau in Fig. \ref{full11},  which gives some idea of the general structure of the potential in this model. 

The early stages of inflation in this model are described by the cascade inflation scenario described in \cite{Kallosh:2017wnt}. Inflation may begin at the upper plateau. Depending on the position on the plateau, the fields either directly moves to smaller values of $\chi$, or first moves towards the blue valley at $\varphi =\sqrt{2} \chi$ (i.e, at $\phi_{2} = 0$), and then moves down along this valley. The process finishes by the second stage of inflation along the deep valley with $\chi =  {2\over \sqrt{6}}\log{\cV_0 \over 8}$ shown in Fig. \ref{2T}, corresponding to the T-model with $\alpha = 2$.

\begin{figure}[t!]
\begin{center}
\includegraphics[scale=0.5]{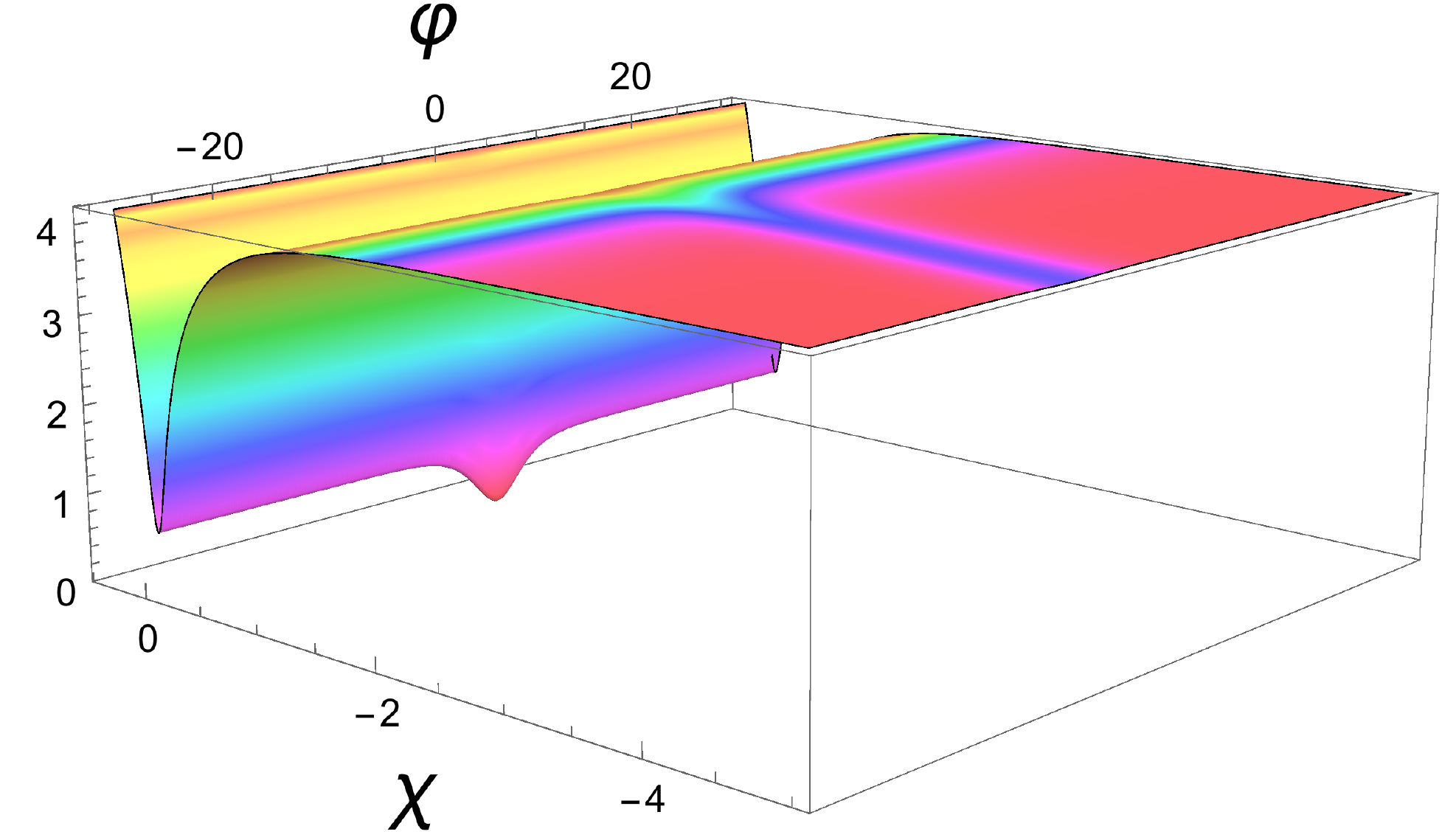}
\end{center}
\caption{\footnotesize Fibre inflation cascades.  Inflation begins at a high dS plateau with the height $M^2 {\cV_0}^4$. Then the fields fall to the narrow valley at $\chi =  {1\over \sqrt{6}}\log{\cV_0^2\over 8}$ to its minimum at $\vp =  {1\over \sqrt{3}}\log{\cV_0^2\over 8}$. In order to simultaneously show the upper plateau, as well as the minimum of the potential  shown in Fig. \ref{2T}, instead of $V$   we plot here $\log(100 V  +1)$ for a particular case $M =m = 0.1$, $\cV_0 = 2\sqrt 2$.}
\label{full11}
\end{figure}

For completeness, one should check whether the inflationary potential is stable with respect to the fields $\vartheta$   and $\psi$ at $\vartheta = \psi = 0$. The calculation is especially simple at the upper plateau shown in Fig. 
\ref{full11}. Indeed, in the limit $\chi \to-\infty$ the potential of the fields $\vartheta$   and $\psi$ is 
\be
V(\vartheta,\psi) = \bigl(m^{2} + M^{2} \cV_0^4\bigr) \, \sec ^2 \frac{\sqrt{2} \vartheta -\psi }{\sqrt{3}}\,  \sec
   \sqrt{\frac{2}{3}} \left(\sqrt{2} \psi +\vartheta  \right) \ .
\ee
By analyzing this expression one finds that the fields $\vartheta$   and $\psi$ on the upper plateau have superheavy masses $m^{2}_{\vartheta} = m^{2}_{\psi} = 2(m^{2} + M^{2} \cV_0^4) = 6H^{2}$, so they are firmly stabilized at $\vartheta = \psi = 0$. 

An investigation of the axion masses along the fibre inflation valley $\chi =  {2\over \sqrt{6}}\log{\cV_0 \over 8}$ is more involved, but it also shows that the fields $\vartheta$   and $\psi$ are stabilized at $\vartheta = \psi = 0$. 

Finally, we show that the masses of scalars at the minimum of the potential are given by
\begin{align}
m_{\tilde\vartheta}^2=4W_0^2,\quad m_{\tilde\psi}^2=\frac12(m^2+8W_0^2),\quad m_{\varphi}^2=\frac{m^2}{6},\quad m_{\chi}^2=12M^2{\cal V}_0^2+\frac{m^2}{3},
\end{align}
where ${\tilde\vartheta}=\frac{1}{\sqrt3}(\vartheta+\sqrt2\psi)$, ${\tilde{\psi}}=\frac{1}{\sqrt{3}}(\sqrt{2}\vartheta-\psi)$ are canonical axions at the minimum.

The potential of this T-model differs from the potential of the original string theory fibre inflation. However, we decided to discuss it there because it has interesting features and it leads to nearly identical observational consequences.

\end{document}